\documentclass[prd,12pt,preprintnumbers,nofootinbib,showpacs]{revtex4}
\usepackage{epsfig,amsmath,amssymb}
\def\KeyWord#1{$\backslash$\IfColor{$\!\!$\textRed{#1}\textBlack}{#1}$\!\!$}
\usepackage{epsfig,amsmath}

\def\Q1{{\bf Q}_1}

\begin{document}
\preprint{\bf FIRENZE-DFF - 421/10/2004}

\title{Pion and kaon condensation in a 3-flavor NJL model}
\author{A. Barducci, R. Casalbuoni, G. Pettini, L. Ravagli}

\affiliation{Department of Physics, University of Florence and
INFN
Sezione di Firenze \\
Via G. Sansone 1, I-50109, Sesto F.no, Firenze, Italy}
\email{barducci@fi.infn.it, casalbuoni@fi.infn.it, 
pettini@fi.infn.it, ravagli@fi.infn.it}

\begin{abstract}                
We analyze the phase diagram of a three-flavor Nambu-Jona-Lasinio
model at finite temperature $T$ and chemical potentials $\mu_u,
\mu_d, \mu_s$. We study the competition of pion and kaon condensation and we propose a physical situation in which kaon condensation could be led only by light quark finite densities.
\end{abstract}
  \maketitle

\section{INTRODUCTION}
\label{int}
In the last few years the study of QCD at finite temperature and densities has been a subject of great interest. In particular, the regime of high temperatures and moderate densities is relevant for the physics of heavy ion collisions: the presence of a tricritical point, suggested in \cite{Barducci:1989wi} is significant to determine the transition from a QGP phase to a hadronic phase \cite{Jakovac:2004pu}. 

Moreover, the presence of an isospin chemical potential, namely of an asymmetry in the densities of light quarks, could complicate this picture: for values of $\mu_I=(\mu_u-\mu_d)/2$ attainable in the experiments, it is expected that the critical lines associated to the light flavors, and therefore the ending points, are mutually shifted. 
The presence of two critical lines would make the transition smoother and would produce observable effects on physical quantities across the transition \cite{Toublan:2003tt,Toublan:2004ks}.

On the other hand, the regime of low temperature and high densities is relevant for compact stars. The existence of stars whose core is so dense that quarks can be treated as fundamental degrees of freedom has been considered for instance in \cite{Steiner:2000bi}: there, the temperature is small enough to safely consider the limit $T\rightarrow0$. 
In this context, the physical scenario is very rich and interesting. For quark densities higher than $300\div400$ MeV, the presence of superconductive phases has been established  \cite{Rajagopal:2000wf}: there, the mass of the strange quark plays the role of distinguishing the pattern of symmetry breaking between a two flavor superconductive (2SC) and a three flavor color flavor locking (CFL) phase, in both cases the gap being of the order of a few tens of MeV. For this reason, an exhaustive study of these phenomena must take into account the strange quark sector.
It is also worth to mention that taking into account weak equilibrium and color and electric neutrality many other interesting superconductive phases could exist \cite{Alford:2000ze,Neumann:2002jm,Alford:2003fq,Shovkovy:2003uu}.

The study of finite density on lattice has well known problems and except for the attempt of Fodor and Katz in ref. \cite{Fodor:2001pe} the only studied case is the one corresponding to low densities \cite{deForcrand:2002ci}.
 For this reason, the only tool available at this moment for the study of the phase diagram in the range from medium to high densities is a class of effective models: the stability of the results obtained with different approaches must be checked to give predictivity to the analyses.

In any case, at zero temperature for values of $\mu_I\geq m_{\pi}/2$ a pion condensed phase should appear, as has been first suggested by \cite{Son:2000xc} in the framework of a chiral model: this phase is confirmed by lattice analyses \cite{Kogut:2002tm}, directly investigable since for $\mu_I\neq0$ the fermion determinant is positive, within chiral models \cite{Splittorff:2000mm} and
within microscopical models \cite{Toublan:2003tt,Klein:2003fy,Frank:2003ve,Barducci:2003un,Barducci:2004tt}. Our previous study in the NJL model \cite{Barducci:2004tt} and that in ref. \cite{Frank:2003ve}, together with the one in random matrix \cite{Klein:2003fy}, showed the effect of both isospin and baryon chemical potentials on the structure of the QCD phase diagram.

By extending this analysis to the third flavor, a kaon condensed phase is expected in the region of high strange quark chemical potential $\mu_s\geq m_K$, as shown by ref \cite{Kogut:2001id}.
The latter analysis being realized in the context of chiral models at zero temperature, 
it would be interesting to study the competition between pion and kaon condensation in the framework of a microscopical model, for a general configuration in the space of thermodynamical parameters $(T,\mu_u,\mu_d,\mu_s)$.

Moreover, it has been suggested a kaon condensed phase in the context of high density nuclear matter \cite{Kaplan:1986yq} and, more recently, in particular in the CFL regime \cite{Bedaque:2001je}: since we don't yet know whether the densities attainable in the core of a neutron star can be high enough to favor a superconductive or a baryon rich phase, we study here the possibility of kaon condensation in a regime where baryons are still present, and where the densities are associated only with non strange quark matter. Also, for the first time we perform the study of kaon condensation by using a microscopical model.

To summarize, in this paper we study pion and kaon condensation at finite temperature and quark densities, by using a three flavor non local NJL model, and we propose a physical scenario in which kaon condensation could be driven only by a light quark finite density. The possibility of a kaon condensation at $\mu_s=0$ could be important in the context of neutron stars, where, by neglecting electroweak effects, the densities are associated only to light quarks.
To this end, we generalize our analyses previously performed in the two flavor sector \cite{Barducci:2004tt,Barducci:2003un}.
In this paper, working at intermediate densities, we do not consider the possibility of diquark condensation, and we turn off the electroweak effects.

\section{The model}
\label{sec:physics}
Our purpose is to explore the general structure of the phase
diagram for chiral symmetry and pion and/or kaon condensation in
three-flavor QCD at non zero quark densities and its evolution in
temperature, by using an effective model with quark degrees of
freedom. To this end, we employ the Nambu-Jona Lasinio model
(NJL) with a form factor included so as to imply a decreasing of
the fermion self-energy at high momenta
\cite{Alford:1997zt}. In this way we generalize, to
three flavors, previous works concerning the simpler two-flavor
case \cite{Barducci:2003un,Barducci:2004tt} which were inspired by ref. \cite{Toublan:2003tt} where the authors had only considered the
case of small differences between the $u$ and $d$ quark chemical
potentials. A complete study of the
two-flavor phase diagram in the space of temperature $T$, baryon
and isospin chemical potentials $\mu_B,~\mu_{I}$ was first done
in the context of a random matrix model \cite{Klein:2003fy}.\\
To our knowledge, so far the only study of meson condensation in
the general case of three-flavor QCD  is based on a chiral
Lagrangian \cite{Kogut:2001id}. Nevertheless the use of a model such as the NJL
model or $ladder$ QCD \cite{Barducci:1988gn} is necessary if we want to
include the effects of a net baryon charge and also study
chiral symmetry restoration through the behaviour of scalar
condensates, besides the pseudoscalar ones. This is not possible
within a chiral Lagrangian approach. As we already stressed in ref. \cite{Barducci:2004tt}, $\it ladder$ QCD and the version of the
NJL model we use in the present work are very similar. Actually
the main difference is that in $\it ladder$ QCD the quark
self-energy depends on the four-momentum whereas in the extended NJL model
it depends only on the three-momentum. This feature greatly
simplifies the numerical calculations and this is the reason why
we employ the NJL model
instead of $\it ladder$ QCD.

Let us thus start with the Lagrangian of the NJL model with three
flavors $u,d,s$, with current masses $m_u=m_d\equiv m$ and $m_s$
and chemical potentials $\mu_u,\mu_d,\mu_s$ respectively

\begin{eqnarray}
{\cal {L}}&=& {\cal {L}}_{0}+{\cal {L}}_{m}+{\cal {L}}_{\mu}+{\cal {L}}_{int}\nonumber\\
&=&{\bar{\Psi}}i{\hat{\partial}}\Psi-{\bar{\Psi}}M\Psi~+~
\Psi^{\dagger}~A~\Psi~+~{G}\sum_{a=0}^{8}\left[\left(
{\bar{\Psi}}\lambda_{a}\Psi
\right)^{2}+\left({\bar{\Psi}}i\gamma_{5}\lambda_{a}\Psi\right)^{2}
\right] \label{eq:njlagr}
\end{eqnarray}
where
\begin{equation} \Psi=\left(
\begin{array}{c} u\\ d\\s
\end{array} \right),~~~~~A=\left(
\begin{array}{c} \mu_u\\0\\0
\end{array} \begin{array}{c} 0\\ \mu_d\\0
\end{array}\begin{array}{c} 0 \\ 0\\ \mu_s
\end{array}\right),
~~~~M=\left(
\begin{array}{c} m\\0\\0
\end{array} \begin{array}{c} 0\\ m\\0
\end{array}\begin{array}{c} 0 \\ 0\\ m_s
\end{array}\right)
\nonumber \end{equation}

\bigskip\noindent $M$ is the quark current mass matrix which is
taken diagonal and $A$ is the matrix of the quark chemical
potentials. As usual $\lambda_0=\sqrt{\displaystyle{{2\over 3}}}~
\mbox{\bf{I}}$ and
$\lambda_{a}~$, $~a=1,...,8~$ are the Gell-Mann matrices.\\
For $M=0$ and $A=0$ or $A\propto \mbox{\bf{I}}$, the Lagrangian is
$U(3)_L\otimes U(3)_R$ invariant. The chiral symmetry is broken by
$M\neq 0$ which also breaks $SU(3)_{V}$ down to $SU(2)_{V}$ as
$m\neq m_s$. However, this symmetry is also lost, since we
generally consider $\mu_u\neq\mu_d$. The remaining symmetry is
thus the product of three independent phase transformations
$U_u(1)\otimes U_d(1)\otimes U_s(1)$.
In this application, we do not consider the 't Hooft determinant, that explicitly breaks $U(1)_A$. Also, we do not take into account the possibility of a di-fermion condensation, which is relevant at low temperatures and high densities, and will be analyzed in a forthcoming publication: for this reason, we limit the validity of our study up to quark chemical potentials of order of $400\simeq500\mbox{MeV}$.

We note that we can express  ${\cal {L}}_{\mu}$ either by using
the variables $\mu_{u},\mu_{d},\mu_{s}$ or, by introducing an
averaged light quark chemical potential
\begin{equation}
\mu_{q}=(\mu_u+\mu_d)/2 \label{eq:muq}
\end{equation}
and the three combinations

\begin{equation}
\mu_B=(\mu_u+\mu_d+\mu_s)/3;~~~\mu_I=(\mu_u-\mu_d)/2;~~~\mu_Y=(\mu_q-\mu_s)/2
\label{eq:defmu}
\end{equation}
which couple to charge densities proportional to the baryon
number, to the third component of isospin and to hypercharge
respectively,

\begin{equation}
{\cal {L}}_{\mu}=\mu_{B}~\Psi^{\dagger}\Psi
~+~\mu_{I}~\Psi^{\dagger}\lambda_{3}\Psi+{2\over\sqrt{3}}~\mu_{Y}~\Psi^{\dagger}\lambda_8\Psi
\label{eq:lmuiso}
\end{equation}

To study whether a pion and/or kaon condensate is formed, we need
to calculate the effective potential. This is obtained by using
the standard technique to introduce bosonic (collective) variables
through the Hubbard-Stratonovich transformation and by integrating
out the fermion fields in the generating functional.\\
The one-loop effective potential we get is:
\begin{equation}
\label{eq:poteff} V=\frac{\Lambda^2}{8G}
(\chi_u^2+\chi_d^2+\chi_s^2+2\rho_{ud}^2+2\rho_{us}^2+2\rho_{ds}^2)+V_{\mbox{log}}
\end{equation}

\begin{eqnarray} \label{eq:vlog}
V_{\mbox{log}}=&& -{1\over\beta}\sum_{n=-\infty}
^{n=+\infty}\int{d^{3}p\over (2\pi)^{3}} ~\mbox{tr log} \left(
\begin{array}{ccc}
h_u & -\gamma_5 F^2(\vec{p})~\Lambda~\rho_{ud}~&-\gamma_5 F^2(\vec{p})~\Lambda~\rho_{us}~\\
\gamma_5 F^2(\vec{p})~\Lambda~\rho_{ud}~ & h_d & -\gamma_5 F^2(\vec{p})~\Lambda~\rho_{ds}~\\
\gamma_5 F^2(\vec{p})~\Lambda~\rho_{us}~& \gamma_5
F^2(\vec{p})~\Lambda~\rho_{ds}~ & h_s
\end{array}
\right)
\nonumber\\
\\
&&~~~~~~~~~h_f=(i\omega_n+\mu_f)\gamma_0~-~\vec{p}\cdot\vec{\gamma}~-~
\left(m_f~+~F^2(\vec{p})~\Lambda~\chi_f\right)\nonumber
\end{eqnarray}
where the form factor $F({\bf
p}^2)=\displaystyle{{\Lambda^2\over \Lambda^2+{\bf p}^2}}$ ($\Lambda$ is a mass scale) has been introduced to mimic asymptotic freedom as in refs. \cite{Alford:1997zt}. In eq. (\ref{eq:vlog}) $\mbox{tr}$ means sum over Dirac, flavor and color indices
and $\omega_{n}$ are the Matsubara frequencies.
The dimensionless fields $\chi_{f}$ and $\rho_{ff^{'}}$ are
connected to the condensates by the following relations

\begin{eqnarray} \label{eq:fields}
\chi_f &=& - ~4G~{\langle{\bar{\Psi}}_f\Psi_f\rangle\over \Lambda}\nonumber\\
\\
\rho_{ff^{'}} &=& -~2G
~{\langle{\bar{\Psi}}_f\gamma_{5}\Psi_{f^{'}}-{\bar{\Psi}}_{f^{'}}\gamma_{5}\Psi_{f}\rangle\over
\Lambda}\nonumber
\end{eqnarray}
and are variationally determined at the absolute minimum of the
effective potential. 
Following the analyses performed within the chiral Lagrangian
approach \cite{Son:2000xc,Kogut:2001id},
we expect a superfluid phase with condensed pions when
the isospin chemical potential exceeds a critical value $\mu_I^C$ ($\mu_I^C=m_{\pi}/2$ at $T=0$); analogously a kaon 
condensation phase is expected when $\mu_Y$, which measures the
unbalance between the light quarks average chemical potential
$\mu_q$ and $\mu_s$, is high enough. For $\mu_u=\mu_d$ and $T=0$ this limit is $\mu_Y^C=m_K/2$. 

To fix the free parameters of the model, which are $\Lambda$, the
average light quark and the strange quark masses $m$, $m_s$ and
the coupling $G=g/\Lambda^2$, we calculate the charged pion and
kaon masses (by fixing  their decay constants to the experimental
values) and the light quark scalar condensate, in the vacuum.

By choosing the free parameters as follows
\begin{equation}
\Lambda=1000~MeV;~~~~~~~g=G~\Lambda^2=6;~~~~~~~~m=1.7~MeV;~~~~m_s=42~MeV~~~~
 \label{eq:parameters}
\end{equation}
we get for the light quark condensate, the
constituent light quark mass, the  pion mass  and the  kaon mass
respectively the results $\langle{\bar
\Psi}_{f}\Psi_{f}\rangle=-(252~MeV)^3$, $M_{q}=385~MeV$,
$m_{\pi}=142MeV$, $m_K=500~MeV$. Furthermore, we find the expected
agreement between $m_{\pi}/2$ and $\mu_I^C$ (within  1$\%$), and also between
$\mu_Y^C$ and $m{_K}/2$ (within  4$\%$).

The very low value we find from our fit for the quark masses is entirely due to
the choice of form factors (instantaneous approximation), see \cite{Schmidt:1994di}, but their ratio $m_s/m\simeq 25$ is in agreement with the value given by the ratio of the pseudoscalar masses.

\section{Phase diagram for chiral symmetry and meson condensation}
\label{sec:physics}
As usual, we determine the phases of the model by looking at the
absolute minimum of the effective potential, given in
Eqs. (\ref{eq:poteff}), (\ref{eq:vlog}). Since there are four
thermodynamical parameters, namely $T,\mu_u,\mu_d,\mu_s$ (or
$T,\mu_B,\mu_I,\mu_Y$), we start from $T=0$ and then follow the
evolution in temperature of two-dimensional slices
in the space of chemical potentials. We recall that the global symmetry in the model is given by the product of three $U(1)$ groups. The possible formation of a non-zero value of one of the pseudoscalar fields $\rho_{ff^{'}}$ implies the spontaneous breaking of a $U(1)$ symmetry with the appearance of a Goldstone boson. The transition to this superfluid phase leaves two $U(1)$ groups unbroken and between them there is always the $U(1)^B_V$. For instance, the formation of a $\rho_{ud}$ condensate implies the breaking of the $U(1)_V^I$ symmetry, whereas $U(1)_V^Y \otimes U(1)_V^B$ are left unbroken. At the threshold one of the charged pions, depending on the sign of $\mu_I$, is the Goldstone boson, whereas the other mesons of the octet remain massive (see also {\cite{Kogut:2001id}).

Thus we characterize different regions of pion or kaon
condensation with the pseudoscalar field which acquires a non
vanishing vacuum expectation value (v.e.v.), whereas the other
v.e.v.'s are vanishing. Then, to facilitate the
physical interpretation and to directly compare it with the results of ref.
\cite{Kogut:2001id}, in the pictures we label these regions
directly with the symbol of the particle that condenses, 
rather than with that of the corresponding field $\rho_{ff^{'}}$.\\
The scalar fields $\chi_{f}$ do not break any symmetry as chiral
symmetry is explicitly broken by the current quark masses. However
these fields undergo cross-over or discontinuous
transitions for given values of the thermodynamical parameters.\\
Thus we distinguish between regions where a non zero v.e.v. of a
scalar field is primarily due to dynamical effects, and we
indicate them with the relative symbol $\chi_{f}$, from regions
where their values are of order $\sim m_{f}/\Lambda$, where we put
no symbol. We also avoid putting the symbol $\chi_f$ in a region where the pseudoscalar field associated with that flavor starts to form, and $\chi_f$ is not yet of order $\sim m_f/\Lambda$.
Furthermore, both for pseudoscalar meson condensation and for the transitions
associated with scalar fields, the boundaries of the regions
separated by discontinuous transitions are represented by solid
lines whereas dashed lines indicate continuous transitions.\\
To start with, we set $\mu_{u}=-\mu_{d}$, which is the case studied by
Toublan and Kogut in ref. \cite{Kogut:2001id}. However, to compare
our results with theirs, we have to remark that we adopt
different combinations of the microscopic chemical potentials
$\mu_{u},\mu_{d},\mu_{s}$. In this case, where $\mu_{q}=0$ (see
Eq. (\ref{eq:muq})), we plot the phase diagram in the plane of
$\mu_{Y}$ and $\mu_{I}$, which are proportional respectively to
$\mu_{s}$ and to $\mu_{u}-\mu_{d}$ and which correspond to half of
the chemical potentials employed in ref. \cite{Kogut:2001id}.\\
In Fig. \ref{fig:Diagmuimus} we plot the quadrant with
$\mu_{Y}>0,\mu_{I}>0$ of the phase diagram at zero temperature.
Let us start by looking at what happens by moving along the vertical
line at $\mu_{I}=0$. In this case one starts from a ``normal"
phase (see also ref. \cite{Kogut:2001id}) characterized by chiral
symmetry breaking by large scalar condensates and then, for
$\mu_{Y}\simeq m_{K}/2\simeq 250~ MeV$, the scalar $\chi_{u}$ and
$\chi_{s}$ condensates smoothly rotate into a $\rho_{us}$
condensate, whereas the $\chi_{d}$ field remains unchanged across
the transition. We have indicated in the picture this new region
with the symbol $\langle K^{+}\rangle$ (rather than with
$\rho_{us}$) as with these signs of the chemical potentials it is
the positively charged kaon to condense. By further increasing
$\mu_{Y}$ the dynamical effect associated with the strange quark
condensation is weakened by a high strangeness density and the
favored phase becomes that with $\chi_{s}\sim m_{s}/\Lambda$ and
large $\chi_{u},\chi_{d}$. The kaon condensate vanishes in this
case through a first order transition. The same situation hereby
described also holds when we move along vertical lines with
$\mu_{I}\lesssim m_{\pi}/2$, with a linear decrease of the
critical $\mu_{Y}$ for kaon condensation and a weak linear
increase for the second transition at higher $\mu_{Y}$.\\
On the other hand, when we move along the horizontal line at
$\mu_{Y}=0$ and we increase $\mu_{I}$, we pass from the ``normal"
phase to a pion condensation phase (in this case the $\pi^{+}$)
described by a finite $\rho_{ud}$, whereas $\chi_{s}$ remains
large. This situation was already known by the previous studies of
the two-flavor NJL model \cite{Barducci:2004tt}. This
transition is also second order. This situation persists at larger
$\mu_{I}$ too. In the intermediate regions there is a competition
between pion and kaon condensation. For instance, by fixing
$\mu_{I}\gtrsim m_{\pi}/2$, starting from low $\mu_{Y}$ 
 in the $\pi^{+}$-condensed phase with a large $\chi_{s}$
and increasing $\mu_{Y}$\footnote{In this case,  working at $\mu_{q}=0$, it simply means to increase $|\mu_{s}|$}, we find that the system
goes to the phase where the $K^{+}$ condenses when $\mu_{Y}\gtrsim m_K/2$.
Consequently $\chi_s$ becomes $\sim m_{s}/\Lambda$ whereas
$\chi_d$ becomes large (if $\mu_{I}<325~MeV$) and the full
transition is discontinuous. When $\mu_{Y}$ is about 300
$MeV$ (with a linear weak increase for growing $\mu_{I}$), the
kaon condensate vanishes and a pion condensed phase shows up,
with all the scalar condensates $\chi_{f}\sim m_{f}/\Lambda$. The
transition is first order. Finally, for high $\mu_{I}$ and $\mu_{Y}$
in the range of values corresponding to kaon condensation, there is a phase analogous to
the previous one, with a kaon condensate instead of a pion condensate
and $\chi_{f}\sim m_{f}/\Lambda$, . This region is
connected to the others through discontinuous transitions
too.

In Fig. \ref{fig:CondT0mui230} we plot the behaviour of the scalar
and pseudoscalar condensates vs. $\mu_{I}$ at $\mu_{q}=0$,
$\mu_{Y}=230~ MeV$ and $T=0$ 
corresponding in Fig. \ref{fig:Diagmuimus} to the path marked by the solid line $a$. It is worth remarking the rotation of the $\chi_{u}$ and $\chi_{s}$
fields into a $\rho_{us}$ field at the continuous transition for
kaon condensation (at $\mu_{I}\simeq 50 MeV$) and the further ''exchange" of the two pseudoscalar fields
$\rho_{us}$ and $\rho_{ud}$ at the pion condensation transition
(at $\mu_{I}\simeq 125 MeV$).

In Fig. \ref{fig:condvert} we plot the behaviour of the scalar and
pseudoscalar condensates vs. $\mu_{Y}$ at $\mu_{q}=0$, $\mu_I=200
MeV$ and $T=0$ (following the path of the solid line $b$ in Fig.
\ref{fig:Diagmuimus}).

In Fig. \ref{fig:Diagallmuimus} the phase diagram of Fig.
\ref{fig:Diagmuimus} is extended to negative values of $\mu_{I}$
and $\mu_{Y}$ too. This picture is thus simply obtained by
reflection of Fig. \ref{fig:Diagmuimus} around its axes at
$\mu_{Y}=0$ and $\mu_{I}=0$. The only difference is that in the
other three quadrants different pseudoscalar mesons of the octet condense
(see also ref. \cite{Kogut:2001id}).

The evolution of the scalar fields for growing temperatures is
well-known. In particular, as expected the growth of
temperature fights against pion and kaon condensation. We find
that the kaon condensed phase disappears at $\mu_{I}=0$ for
$T>85MeV$. In Fig. \ref{fig:DiagmuimusT100} we show the evolution
of the phase diagram of Fig. \ref{fig:Diagmuimus} at $T=100~MeV$.
The region for kaon condensation has shrunk and the transitions
associated with chiral symmetry approximate restoration have become
continuous. Above $T\simeq 110~MeV$ the regions of kaon
condensation disappear. In Fig. \ref{fig:DiagmuimusT140} we plot
the further evolution of the phase diagram of Fig.
\ref{fig:Diagmuimus} at $T=140~MeV$. In the range of values of
$\mu_{I},\mu_{Y}$ that we have considered, only the
regions of pion condensation and the usual regions with chiral
symmetry breaking separated by continuous transitions from
analogous phases with $\chi_{s}\sim m_{s}/\Lambda$ for large
$\mu_{Y}$ remain. Finally, pion condensation disappears above the
cross-over critical temperature $T=192~MeV$.\\

\section{Kaon condensation in the region of high $\mu_q$}

So far we have considered $\mu_{u}=-\mu_{d}$ (and thus $\mu_q=0$). Besides this case,
the relevant features of the full phase diagram can be grasped by
considering two other possibilities, which are $\mu_{u}=\mu_{d}$
(and thus $\mu_{I}=0$), and $\mu_{s}=0$, which has been 
already considered in ref. \cite{Barducci:2004tt}.
In this section, we suggest a hypothesis for a kaon condensation driven mainly by light quark finite densities, with $\mu_s$ low or even zero. We recall that in the case of $\mu_s=0$ there is an equal number of strange quarks and anti-quarks, and in that case kaon condensation 
would concern a light particle (antiparticle), associated to the external field,  and a strange anti-quark (quark) belonging to the sea.
In ref. \cite{Barducci:2004tt}, we had not inserted the pseudoscalar fields with strangeness content: the possibility of kaon condensation at $\mu_s=0$ would modify the phase diagram in the ($\mu_u,\mu_d$) plane of ref. \cite{Barducci:2004tt} with the insertion of regions with condensed kaons. 
However, except for this, no further modification
would be necessary, whereas the possibility of kaon condensation
at zero or low values of $\mu_{s}$ can be also studied for $\mu_{I}=0$,
which is therefore the last situation we examine.\\
From its definition in Eq. (\ref{eq:defmu}), we see that $\mu_{Y}$
measures the difference between the strange quark chemical
potential and $\mu_{q}$ (which at $\mu_{I}=0$ is the chemical
potential associated with one of the light flavors) similarly to
$\mu_{I}$ which measures the unbalance between the $u$ and $d$
quark chemical potentials. Therefore as pion condensation
can occur, at $T=0$ and for instance at $\mu_{d}=0$, when $\mu_{u}\gtrsim
m_{\pi}$, similarly we could expect a kaon condensed phase at
$T=0$ and $\mu_{s}=0$ when $\mu_{q}\gtrsim m_{K}$. However, within
our approximations, this does not happen, the reason being that
$m_{K}$ is higher than the critical value of $\mu_{q}$ for the
melting of the dynamical part of $\langle {\bar
\Psi}_{f}\Psi_{f}\rangle$, $f=u,d$. In other words, when the
condensates of the light quarks start to decrease to values $\sim
m/\Lambda$, the effective kaon mass starts to increase and kaon
condensation always remains unfavored (the same does not happen at
$\mu_{q}=0$ because the melting of the dynamical part of $\langle
{\bar s}s\rangle$ occurs at higher values of $\mu_{s}$ and actually we find
kaon condensation as shown in the last section).
Nevertheless, since we do not have taken into account the effects
of di-quark condensates which could break chiral symmetry, it
would be interesting to further analyze what happens with their
inclusion. We will analyze this case in further work.

For the time being, let us show, in Fig. \ref{fig:Diagmuqmus},
the $T=0$ phase diagram in the $(\mu_{q},\mu_{s})$ plane at
$\mu_{I}=0^{-}$, where $0^{-}$ stands for an infinitesimally small
negative value, which is necessary to decide which are the kaons
that condense (in any case, within a neutron star a finite negative isospin chemical potential is present).
 The structure of this phase diagrams recalls that
of \cite{Barducci:2004tt} for pion condensation in the
$(\mu_{u},\mu_{d})$ plane, with a ``normal" phase in the middle,
characterized by high values of the scalar condensates
$\chi_{u},\chi_{d},\chi_{s}$. This phase is separated from the regions of chiral symmetry approximate restoration (relative to one or more flavors depending on the values of chemical potentials),
and is characterized by the respective scalar field $\chi_{f}\sim
m_{f}/\Lambda$. Furthermore this region is separated through discontinuous transitions,
both from the ``normal" phase and from the regions of kaon
condensation. There are four different regions of this kind,
distinguishable by the condensing kaon and by the values of the associated scalar condensates. Finally, the ``normal" phase turns
into a region of kaon condensation through the smooth rotation of
$\chi_{s}$ and one of the light flavors (according to the sign of
$\mu_{Y}$ one between $\chi_{u}$ or $\chi_{d}$ remains large) into
a kaon condensate, when $\mu_{Y}\gtrsim m_{K}/2$, with a
second order phase transition. We can see that the presence of a high value of $|\mu_q|$
 lowers the corresponding $|\mu_s|$ necessary to have kaon condensation: moreover, a value of $\mu_I\neq 0$ would favor a kaon condensation with lower values of $|\mu_q|$ and/or $|\mu_s|$.\\
Since we do not know exactly the critical value $\mu_q^C$ for chiral symmetry restoration, and since we are interested here in the case where $\mu_q^C>m_{K}$, instead of varying $\mu_q^C$ (for instance performing a different fit) we take the kaon mass as a free parameter, and we present, in Fig. \ref{fig:limsuk}, the same result as in Fig.
\ref{fig:Diagmuqmus}, for $m_{K}=300~MeV$, which is
the upper limit for having a kaon condensation at
$\mu_{s}=0$.\\

\section{Conclusions}
In this paper we have presented a calculation of the QCD phase diagram obtained by using a three-flavor NJL model. We 
have considered the plane ($\mu_I,\mu_Y$) to study the competition of kaon and pion condensation at zero and finite temperature and to establish a comparison with a previous analysis performed in a chiral model \cite{Kogut:2001id}. The results of the microscopical model accurately reproduce
 those achieved in the effective model only for low densities, since in the high strangeness density regime approximate chiral symmetry restoration (relative to the strange sector) occurs, disfavoring kaon with respect to pion condensation.
Moreover we have suggested the possibility of kaon condensation driven only by light quark densities: this situation could be of interest in the core of compact stars, where the densities are approximatively associated only to light flavors. Up to now, we have neglected the possibility of di-quark condensation: however, we are currently working on an extension of this analysis that takes into account the superconductive phases too, so as to
reproduce the physical picture proposed in the context of effective models for the K0-CFL phase \cite{Bedaque:2001je},
 and to understand the behaviour of color superconductivity by varying arbitrarily the three quark chemical potentials \cite{Bedaque:1999nu}. Eventually, the inclusion of the $\beta$-equilibrium as well as charge and color neutrality, will give us a reasonable picture of cold quark matter within stars, and will make possible a comparison with experimental data.



\eject

\begin{figure}[htbp]
\begin{center}
\includegraphics[width=10cm]{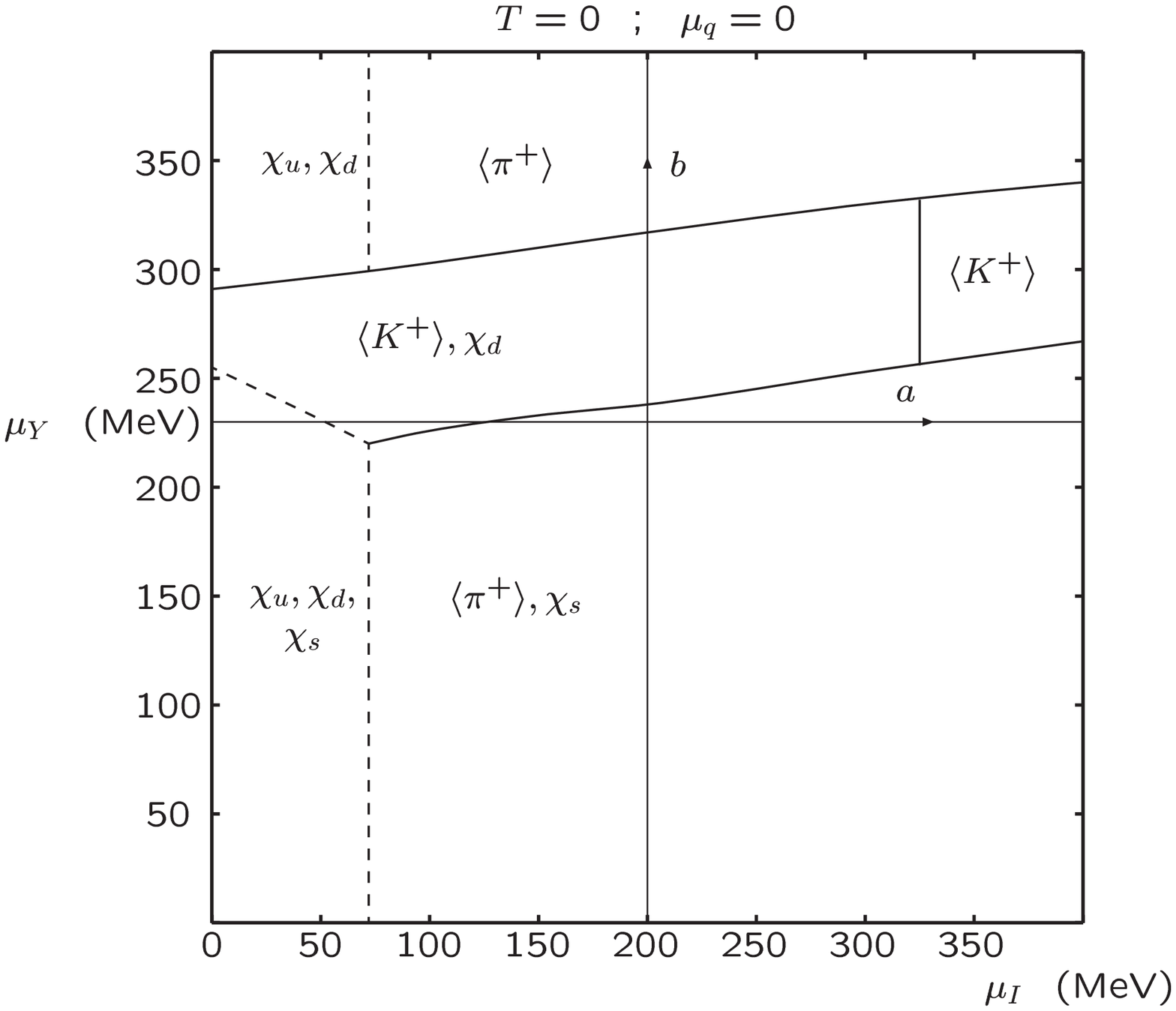}
\end{center}
\caption{\it Phase diagram for chiral symmetry restoration and meson condensation in the plane ($\mu_I,\mu_Y$) at $\mu_q=0$ and $T=0$. Different regions are specified by the non vanishing of a given condensate, whereas the others are vanishing ($\rho_{ud},\rho_{us}$) or order $\sim m_f/\Lambda$ ($\chi_u,\chi_d,\chi_s$). Dashed lines are for the continous vanishing of pseudoscalar fields, whereas solid lines are for discontinuous behaviours. The solid lines
a and b refer to specific paths at fixed values of $\mu_Y=230~\mbox{MeV}$ (line a) and $\mu_I=200~\mbox{MeV}$ (line b).} \label{fig:Diagmuimus}
\end{figure}

\begin{figure}[htbp]
\begin{center}
\includegraphics[width=10cm]{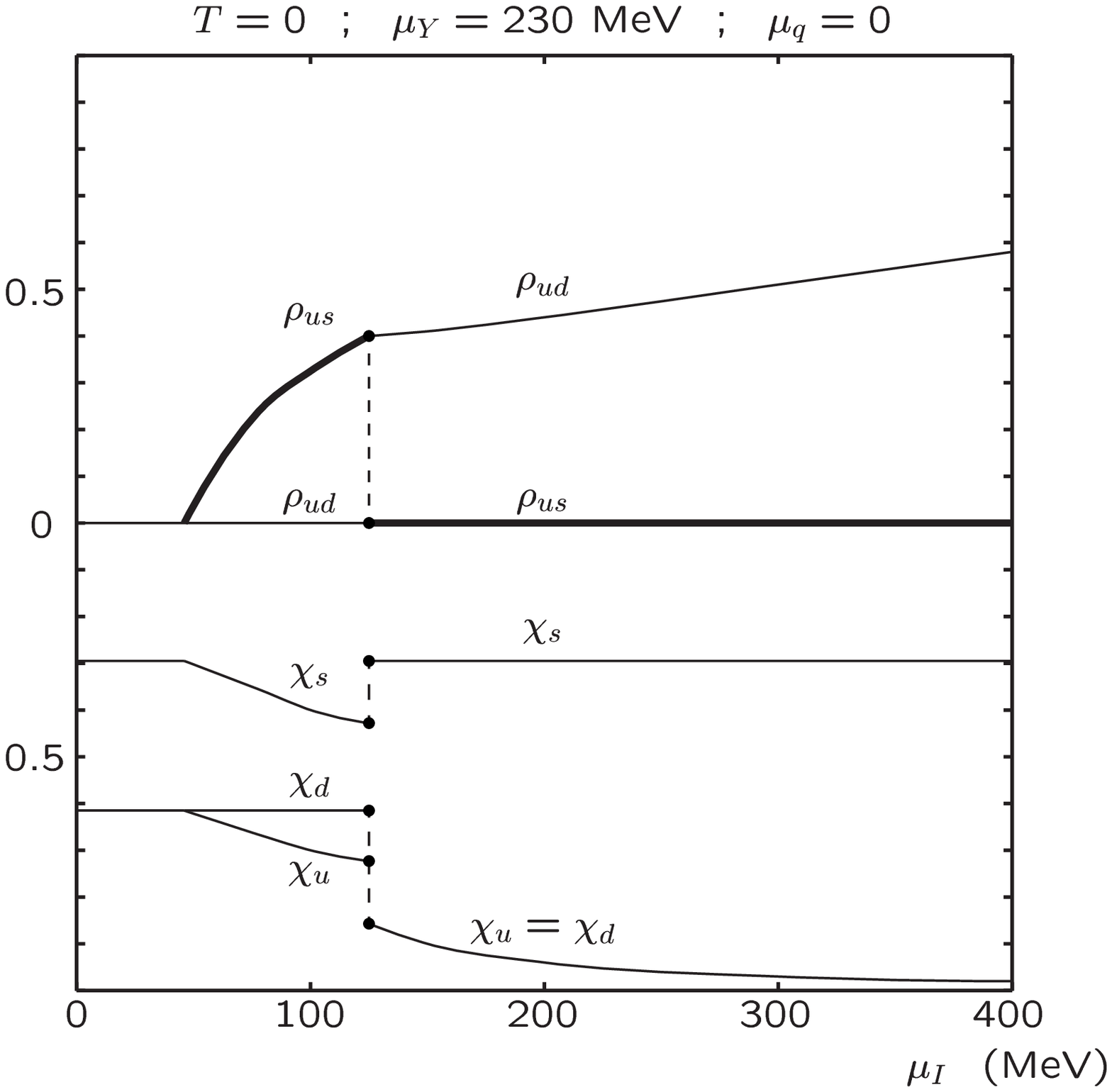}
\caption{\it Scalar and pseudoscalar condensates vs. $\mu_I$, for $T=0$, $\mu_q=0$ and $\mu_Y=230~{MeV}$. The path followed in the phase diagram of Fig. \ref{fig:Diagmuimus} is that of the solid line a.} \label{fig:CondT0mui230}
\end{center}
\end{figure}

\begin{figure}[htbp]
\begin{center}
\includegraphics[width=10cm]{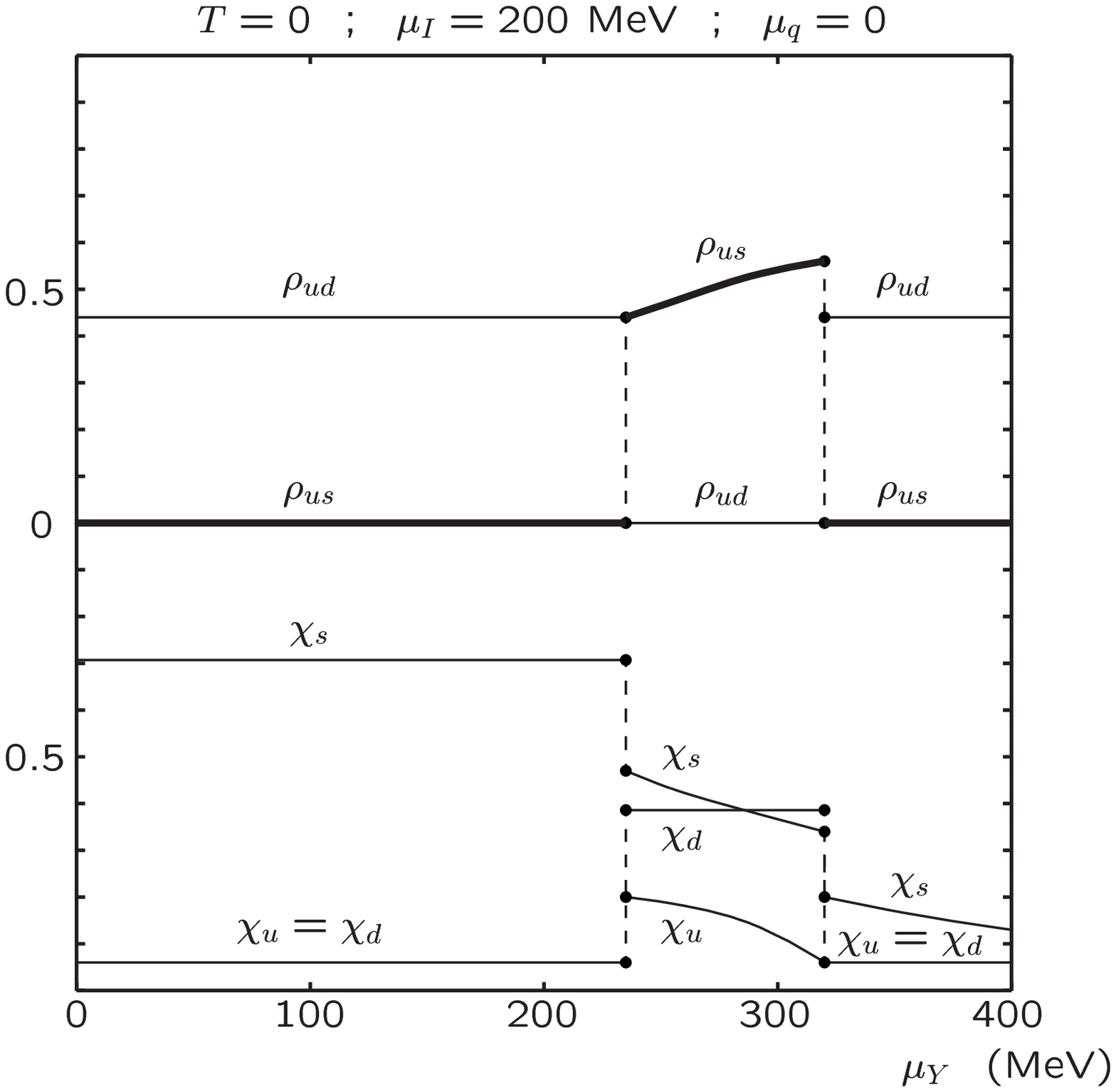}
\caption{\it Scalar and pseudoscalar condensates vs. $\mu_Y$, for $T=0$, $\mu_q=0$ and $\mu_I=200~{MeV}$. The path followed in the phase diagram of Fig. \ref{fig:Diagmuimus} is that of the solid line b.} \label{fig:condvert}
\end{center}
\end{figure}

\begin{figure}[htbp]
\begin{center}
\includegraphics[width=10cm]{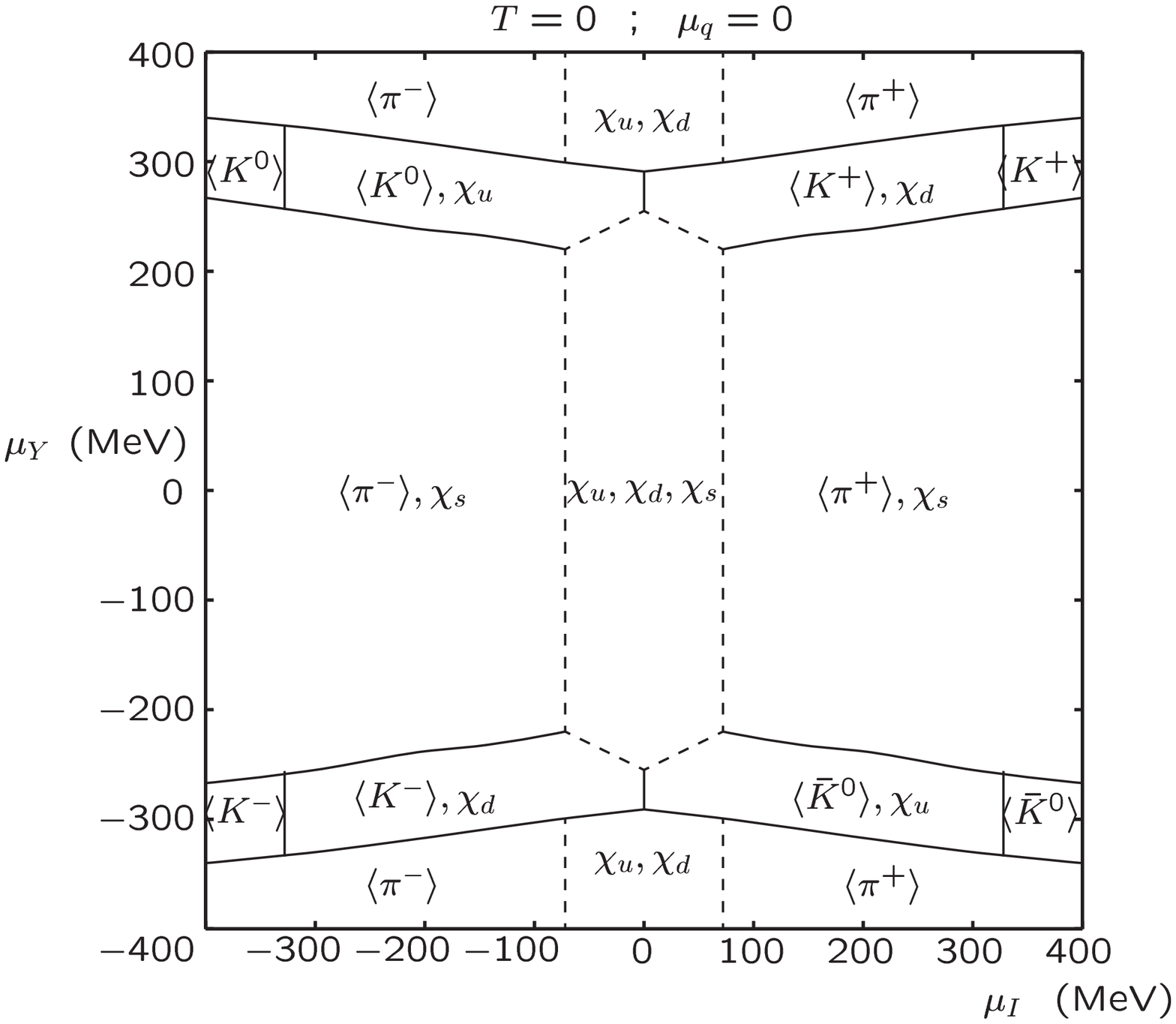}
\end{center}
\caption{\it Phase diagram for chiral symmetry restoration and meson condensation in the plane ($\mu_I,\mu_Y$) at $\mu_q=0$ and $T=0$. Different regions are specified by the non vanishing of a given condensate, whereas the others are vanishing ($\rho_{ud},\rho_{us},\rho_{ds}$) or order $\sim m_f/\Lambda$ ($\chi_u,\chi_d,\chi_s$). Dashed lines are for the continous vanishing of pseudoscalar fields, whereas solid lines are for discontinuous behaviours.} \label{fig:Diagallmuimus}
\end{figure}

\begin{figure}[htbp]
\begin{center}
\includegraphics[width=10cm]{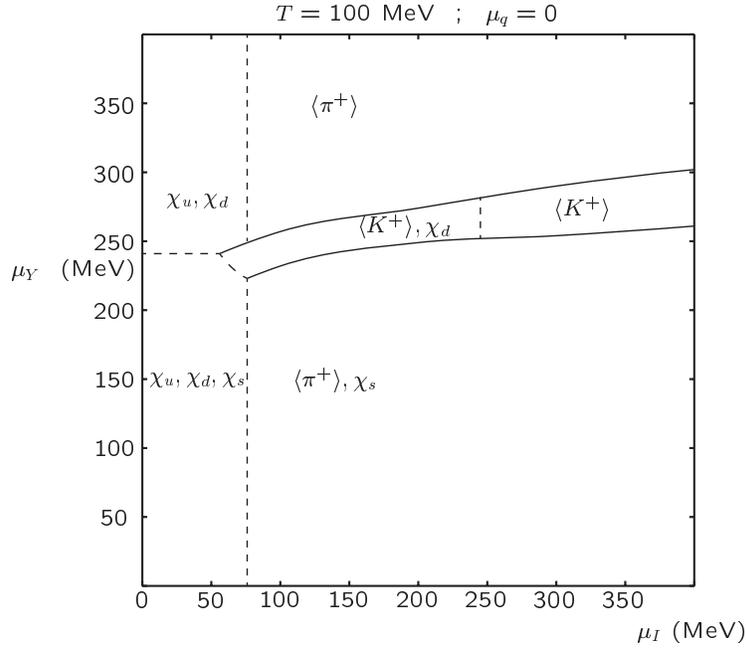}
\caption{\it Phase diagram for chiral symmetry restoration and meson condensation in the plane ($\mu_I,\mu_Y$) at $\mu_q=0$ and $T=100~MeV$, which is above the temperature of the tricritical point. Different regions are specified by the non vanishing of a given condensate, whereas the others are vanishing ($\rho_{ud},\rho_{us}$) or order $\sim m_f/\Lambda$ ($\chi_u,\chi_d,\chi_s$). Dashed lines are for the continous vanishing of pseudoscalar fields or for cross-over transitions for scalar fieds, whereas solid lines are for discontinuous behaviours.} \label{fig:DiagmuimusT100}
\end{center}

\end{figure}

\begin{figure}[htbp]
\begin{center}
\includegraphics[width=10cm]{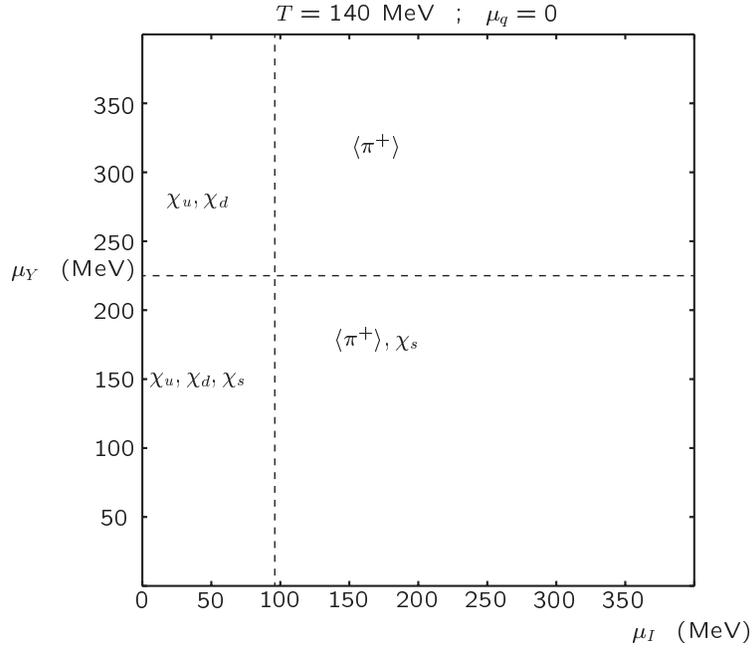}
\caption{\it Phase diagram for chiral symmetry restoration and pion condensation in the plane ($\mu_I,\mu_Y$) at $\mu_q=0$ and $T=140~MeV$, which is above the highest temperature $\sim~110~MeV$ to have kaon condensation. Different regions are specified by the non vanishing of a given condensate, whereas the others are vanishing ($\rho_{ud}$) or order $\sim m_f/\Lambda$ ($\chi_u,\chi_d,\chi_s$). Dashed lines are for the continous vanishing of pseudoscalar fields or for cross-over transitions for scalar fieds.} \label{fig:DiagmuimusT140}
\end{center}

\end{figure}

\begin{figure}[htbp]
\begin{center}
\includegraphics[width=10cm]{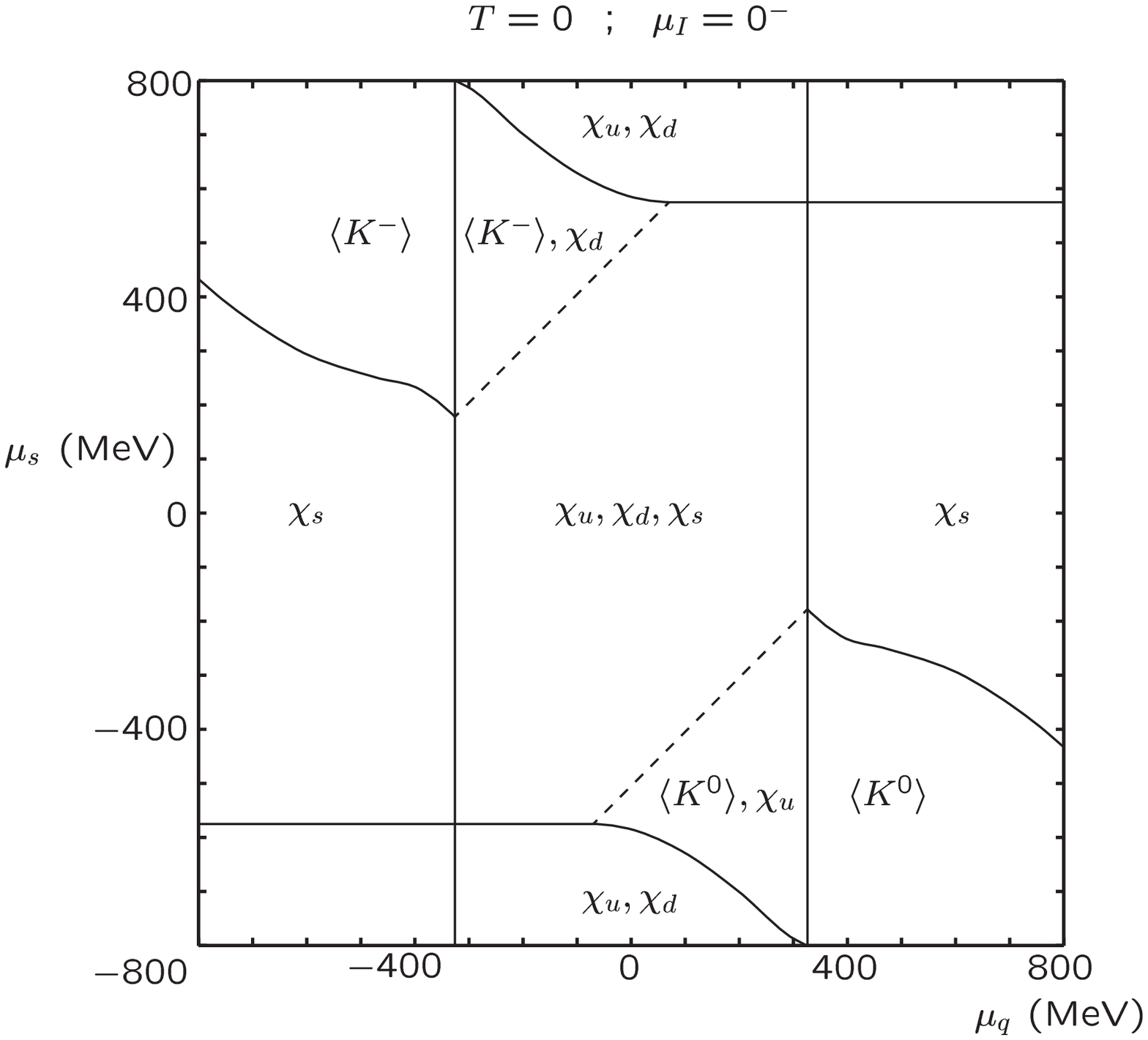}
\end{center}
\caption{\it Phase diagram for chiral symmetry restoration and kaon condensation in the plane ($\mu_q,\mu_s$) at $\mu_I=0^-$ and $T=0$. Different regions are specified by the non vanishing of a given condensate, whereas the others are vanishing ($\rho_{us},\rho_{ds}$) or order $\sim m_f/\Lambda$ ($\chi_u,\chi_d,\chi_s$). Dashed lines are for the continous vanishing of pseudoscalar fields, whereas solid lines are for discontinuous behaviours.}\label{fig:Diagmuqmus}
\end{figure}

\begin{figure}[htbp]
\begin{center}
\includegraphics[width=10cm]{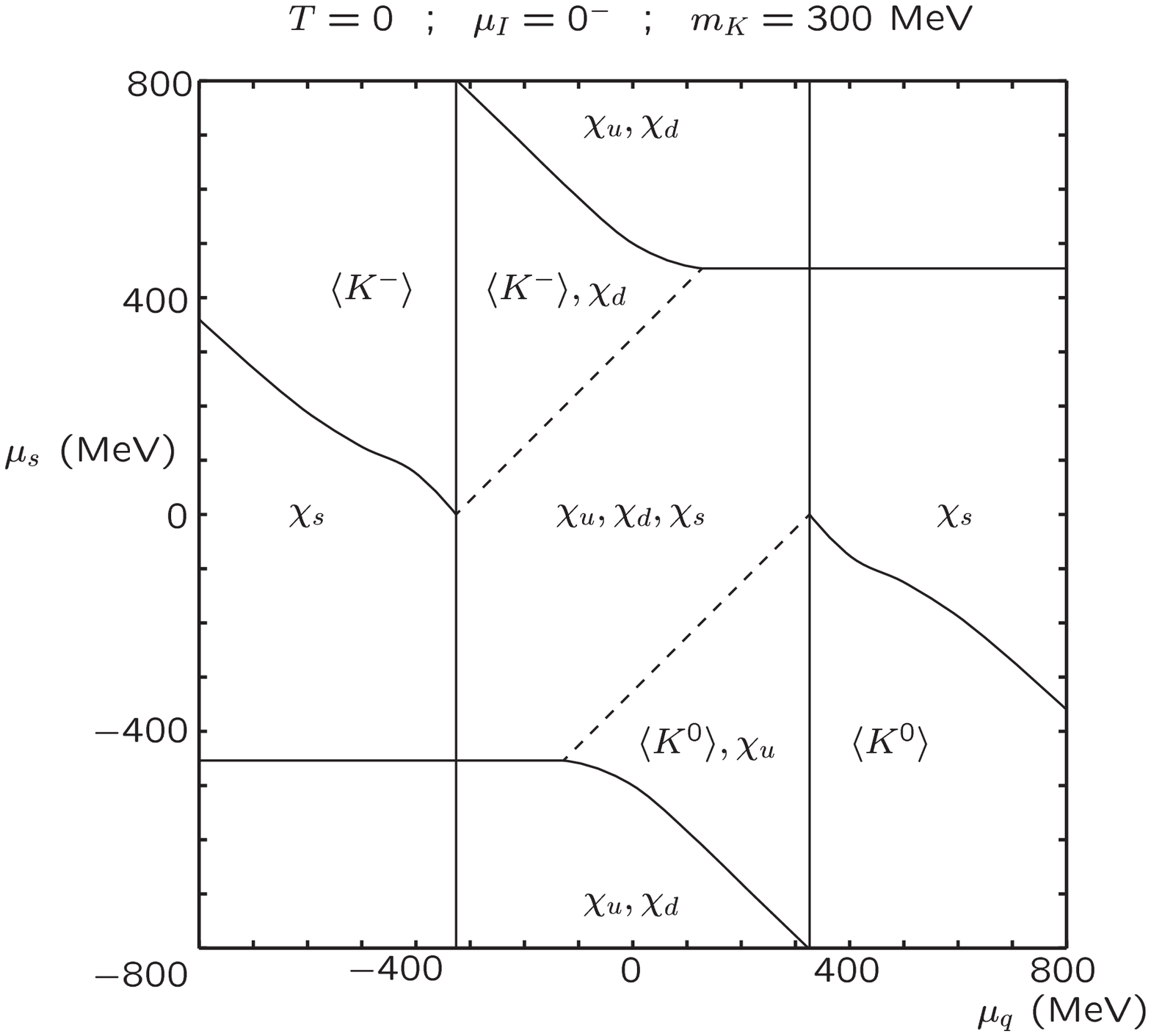}
\end{center}
\caption{\it Phase diagram for chiral symmetry restoration and kaon condensation in the plane ($\mu_q,\mu_s$) at $\mu_I=0^-$ and $T=0$, and for $m_K=300~MeV$. Different regions are specified by the non vanishing of a given condensate, whereas the others are vanishing ($\rho_{us},\rho_{ds}$) or order $\sim m_f/\Lambda$ ($\chi_u,\chi_d,\chi_s$). Dashed lines are for the continous vanishing of pseudoscalar fields, whereas solid lines are for discontinuous behaviours.}\label{fig:limsuk}
\end{figure}

\end{document}